This is the accepted version of the manuscript.

Publisher version:

# Mechanoreceptive Aβ primary afferents discriminate naturalistic social touch inputs at a functionally relevant time scale

Shan Xu, Steven C. Hauser, Saad S. Nagi, James A. Jablonski, Merat Rezaei, Ewa Jarocka, Andrew G. Marshall, Håkan Olausson, Sarah McIntyre, Gregory J. Gerling

*Abstract*— Interpersonal touch is an important channel of social emotional interaction. How these physical skin-to-skin touch expressions are processed in the peripheral nervous system is not well understood. From microneurography recordings in humans, we evaluated the capacity of six subtypes of cutaneous mechanoreceptive afferents to differentiate human-delivered social touch expressions. Leveraging statistical and classification analyses, we found that single units of multiple mechanoreceptive Aβ subtypes, especially slowly adapting type II (SA-II) and fast adapting hair follicle afferents (HFA), can reliably differentiate social touch expressions at accuracies similar to human recognition. We then identified the most informative firing patterns of SA-II and HFA afferents, which indicate that average durations of 3-4 s of firing provide sufficient discriminative information. Those two subtypes also exhibit robust tolerance to spike-timing shifts of up to 10-20 ms, varying with touch expressions due to their specific firing properties. Greater shifts in spike-timing, however, can change a firing pattern's envelope to resemble that of another expression and drastically compromise an afferent's discrimination capacity. Altogether, the findings indicate that SA-II and HFA afferents differentiate the skin contact of social touch at time scales relevant for such interactions, which are 1-2 orders of magnitude longer than those for non-social touch.

## I. INTRODUCTION

TOUCH is an often used medium for facilitating social relationships and emotional interactions. For example, one might lightly tap another person to get their attention, or stroke a partner's arm to offer a sense of calm. Between people in close relationships, and even between strangers, many social touch expressions are intuitively understood [1]–[4]. The appreciation of emotion is commonly thought to be a centrally mediated process performed by frontal and temporal brain structures that integrate a multitude of peripheral and cross-cortical sensory information [5]. However, the peripheral nervous system may already be organized to facilitate the selection and processing of potentially socially relevant stimuli [6]. Reliable signaling from peripheral afferents could form the basis of the somatosensory and affective perception in the central nervous system. In our evolutionary history, such peripheral encoding may also have acted as scaffolding for the development of cross-sensory, cortical processing of emotion [7].

Among peripheral tactile afferents, percepts tied to social and emotional touch are thought to be influenced prominently by C-tactile (CT) afferents [8], [9]. These afferents can be preferentially activated by light stroking contact at 1-10 cm/s velocities [8] with temperatures similar to human skin [10]. Their firing frequencies have been found to be proportionally correlated with subjectively perceived pleasantness [8], [10], and both follow an inverted U-shape curve along with stroking velocity with a peak around 1-10 cm/s. Such an inverted U-shape relationship between pleasant sensation and stimuli velocity has been widely and reliably reproduced on the population level [11]–[14], and has been suggested to be related with the firing patterns of CT afferents. Recent work has, however, encountered difficulty in reproducing such trends among individual participants [15], which suggests a more complex view of pleasantness and affective touch and a plausible role of other afferent types. Meanwhile, the firing properties of CT afferents have mainly been characterized in response to controlled stimuli [8]–[10], [15], such as rotary actuated brushing, while less explored under naturalistic, human-to-human touch.

In contrast to CT afferents, low-threshold mechanosensitive (LTM) Aβ afferents have been investigated in a wider variety of scenarios, especially in relation to discriminative touch such as surface roughness perception or skin-object friction. Pre-defined, well-controlled mechanical stimuli have been used to decouple and examine stimulus attributes, one at a time [16]–[19]. Across these studies, different tactile cues, e.g., pressure, vibration, shape, texture, the deflection of hair follicles, etc., were shown to be mainly encoded by certain Aβ subtypes [16]–[21]. Moreover, the perception of certain elementary cues has

This work was supported in part by the National Science Foundation under Grant IIS-1908115, National Institutes of Health under Grant NINDS R01NS105241, and Swedish Research Council under Grant 2020-01085. *(Corresponding authors: Sarah McIntyre, Gregory J. Gerling).* Shan Xu, Steven C. Hauser, and Saad S. Nagi contributed equally.

Shan Xu, Steven C. Hauser, James A. Jablonski, Merat Rezaei, and Gregory J. Gerling are with the University of Virginia, Charlottesville, VA 22903 USA. E-mail: {sx3ky, sch5zc, jaj4zcf, mr3wq, gg7h}@virginia.edu.

Saad S. Nagi, Håkan Olausson, and Sarah McIntyre are with Linköping University, Linköping, Sweden. E-mail: {saad.nagi, hakan.olausson, sarah.mcintyre}@liu.se.

Ewa Jarocka is with Linköping University, Linköping, Sweden, and Umeå University, Umeå, Sweden. E-mail: ewa.jarocka@liu.se

Andrew G. Marshall is with University of Liverpool, Liverpool, United Kingdom. E-mail: andrew.marshall@liverpool.ac.uk

been invoked via the intraneural electrostimulation, e.g., slowly adapting type I and fast adapting units for pressure and flutter/vibration [22]–[24]. However, device-delivered stimuli do not reflect the full range of naturalistic touch we encounter in everyday life. Indeed, in discriminative touch scenarios that invoke multiple tactile cues, e.g., object manipulation [25] and natural textures [26], single Aβ subtypes provide overlapping and complementary information [27]. As multiple tactile cues vary simultaneously in human-to-human touch [1], [4], the analysis of their firing patterns becomes more difficult.

Here, we investigated how the spike firing patterns of Aβ and CT human peripheral afferents encode information about the mechanical inputs produced by human-delivered social touch expressions. Microneurography experiments were conducted to record from single unit, peripheral afferents in human participants with natural human touch as the stimulus. Six standardized social touch expressions are delivered, each composed of complex dynamic skin mechanical properties but with socially distinct meanings. We first characterized afferents' firing properties, i.e., firing frequency and number of spikes, for comparison to prior studies with well-controlled mechanical contact. Then, machine-learning classifiers were developed to examine the capability of each afferent subtype in differentiating the expressions, for comparison with perceptual studies. Two classification strategies were employed following the theory of temporal coding and rate coding of the neural firing [28], respectively. Moreover, with these models, we evaluated the classification performance of different segments of the neural recordings and their sensitivity to spike-timing noise, to identify the most informative firing patterns for each expression. Overall, the encoding performance of peripheral afferents and their firing properties in human-delivered social touch shed light on the information present at the periphery, which may affect the strategies available to the central nervous system for processing social intent, emotional state or affiliative alignment from physical skin contact.

## II. EXPERIMENTAL METHODS

### A. Participants - Touch Receivers

Twenty healthy participants (23-35 years old with one exception of 50 years old) were recruited through local advertisement and a mailing list. All participants provided informed consent in writing before the experiment. The study was approved by the Swedish Ethical Review Authority (Dnr 2017/485-31) and complied with the revised Declaration of Helsinki.

### B. Standardized Touch Expressions

Based on social touch communication between people in a close relationship, we used a previously developed set of six, standardized social touch expressions, including "attention," "happiness," "calming," "love," "gratitude" and "sadness" [4], [29]. Those expressions have been validated to be reliably and effectively recognizable by naïve stranger participants with accuracy similar or even higher than people with close relationships [4]. Experimenters were trained to deliver those standardized social touch expressions in the same way as in our preceding studies [4]. Since during microneurography, it is logistically difficult to simultaneously obtain direct psychological responses from participants, such as their subjective emotional perception, we connected the emotional meanings of those social touch expressions though this standardized expression set.

More specifically, the touch expression of "attention" comprised 4 bursts of 4-5 repetitive taps with the index finger, each burst lasting approximately 1.5 s, with approximately 1 s between. "Happiness" consisted of continuous random playful tapping using multiple fingers, and moving up and down the arm. "Gratitude" consisted of patting (3-4 pats with multiple fingers, lasting approximately 2 s) alternated with holding (long grasp with the whole hand, lasting approximately 2 s). "Calming" involved 4 repeated slow strokes down the arm with the whole hand, each lasting approximately 2 s, with approximately 0.5 s between. "Love" involved a continuous back-and-forth light stroking with the fingertips up and down the arm. Finally, "sadness" consisted of a sustained hold with firm but gentle squeezing.

### C. Microneurography

Neural recordings were performed with equipment purpose-built for human microneurography studies from ADInstruments (Oxford, UK; setup 1) or the Physiology Section, Department of Integrative Medical Biology, Umeå University (setup 2). The course of the radial nerve just above the elbow was visualized using ultrasound (LOGIQ e, GE Healthcare, Chicago, IL, USA). A high-impedance tungsten recording electrode was inserted percutaneously and with ultrasound guidance it was inserted into the nerve. Where needed, weak electrical stimuli through that electrode were delivered to localize the nerve (0.02-1 mA, 0.2 ms, 1 Hz; FHC, Inc. Bowdoin, ME, USA). The electrode was insulated, except for the ~5 μm bare tip, with a typical length of 40 mm and shaft diameter of 0.2 mm. In addition to the recording electrode, an indifferent (uninsulated) electrode was inserted subcutaneously, approximately 5 cm away from the nerve. Once the electrode tip was intra-fascicular, minute movements were made to the recording electrode, manually or with a pair of forceps until a single afferent signal was isolated.

Each low-threshold mechanosensitive cutaneous afferent (all soft-brush sensitive) was classified by its physiological characteristics, as per the criteria used in [30], [31]. Briefly, individual Aβ low-threshold mechanoreceptors were separated into fast and slowly adapting types based on their adaptive responses to ramp-and-hold indentation of the skin. Three groups of fast adapting units were identified as follows: fast adapting hair follicle (HFA), responsive to hair deflection and light air puffs; fast adapting Pacinian corpuscle (FA-II), comprising a single spot of maximal sensitivity and robust response to remote tapping; fast adapting field (Field), comprising multiple spots of high sensitivity with no response to hair displacement or remote tapping of the skin. Two groups of slowly adapting units, i.e., type I (SA-II) and type II (SA-II), were identified where several features were examined including spontaneous firing, stretch sensitivity, and receptive field characteristics. In addition, an inter-spike interval pattern to sustained indentation (100 mN for 30 s) was tested. Coefficients of variation of inter-spike intervals for all SA-IIs (4 units) were in the range of 0.15 to 0.23. This was also measured for one SA-I and its coefficient of variation was 1.92. These values are consistent with previous observations [23], [30], [32]. Single muscle spindle (MS) afferents were identified by stretch of the receptor-bearing muscle along its line of action. These were not further classified into primary and secondary afferents.

Mechanical thresholds of all cutaneous afferent fibers were measured using Semmes-Weinstein monofilaments (nylon fiber; Aesthesio, Bioseb, Pinellas Park, FL, USA), except HFA whose preferred stimulus is hair movement so responses to light air puffs were determined. The monofilaments were applied manually with a rapid onset until the monofilament buckled: If a unit responded to the same (weakest) monofilament in at least 50% of trials, it was taken as the mechanical threshold. Based on prior work showing that 4 mN threshold divides the low threshold (<4 mN) and high threshold (≥ 4 mN) cutaneous afferent populations in hairy skin [23], [31], only those afferents with thresholds below 4 mN were considered. Further, any cutaneous afferent with a receptive field located at a site inaccessible for the delivery of expressions was discarded.

All neural data were recorded and processed using LabChart Pro for setup 1 (v8.1.5 and PowerLab 16/35 hardware PL3516/P, ADInstruments, Oxford, UK) and SC/ZOOM for setup 2 (Physiology Section, Department of Integrative Medical Biology, Umeå University). Action potentials were distinguished from background noise with a signal-to-noise ratio of at least 2:1 and were confirmed to have originated from the recorded afferent by a semi-automatic inspection of their morphology. For further details see [23].

### D. Experimental Procedure

Participants were seated in a comfortable chair and pillows were provided to ensure maximal comfort. Designed standardized expressions were applied by trained experimenters to the receptive field of single neurons during microneurography recordings. The experimenter received audible spoken cues, first the cue-word, then a countdown (3, 2, 1, go). They were instructed to perform the touch starting from the "go"-signal until they heard a stop signal (3, 2, 1, stop), creating a continuous time window of touch for 10 s. Those cues could be heard by both experimenters and participants, but would not influence microneurography recordings, since peripheral tactile afferents do not receive top-down projections from higher-order neurons. The experimenter was first familiarized with the afferent's receptive field and was asked to touch an area of skin including but not limited to the receptive field (Fig. 1A). They were also required not to perform vigorous movements to avoid dislodging the recording electrode.

We recorded 41 low-threshold primary afferent units in total (Fig. 1B), which were classified into seven subtypes: Field (5 units), HFA (7 units), FA-II (5 units), SA-I (2 units), SA-II (4 units), CT (6 units), and MS (12 units). Among them, Field, HFA, FA-II, SA-I, and SA-II were classified as Aβ afferents.

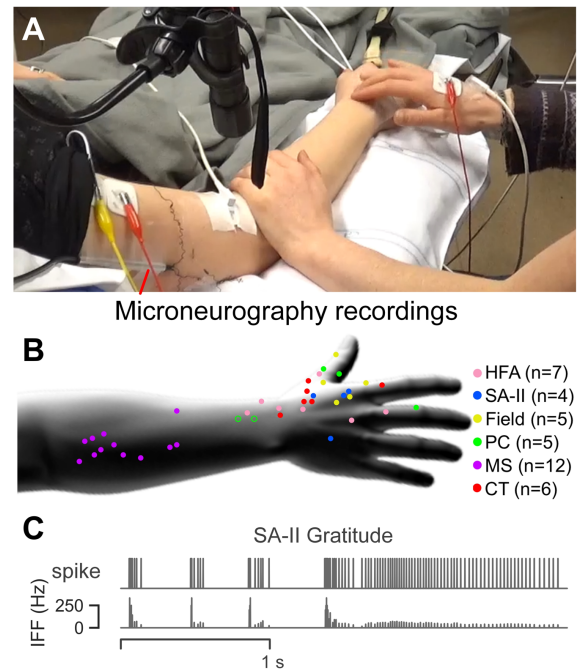

**Fig. 1.** Experimental setup for microneurography experiments. (A) Standardized touch expressions were delivered over receptive fields of identified afferents by trained experimenters. Microneurography recordings were collected from the upper arm. (B) Multiple units were recorded for each of the six afferent subtypes. For cutaneous afferents, each dot represents the location of an individual receptive field. For two FA-II afferents in the forearm (open circles), the precise location of the receptive field was not documented. For muscle afferents, the dots are shown simply to illustrate where the gestures were delivered. The n-value denotes the number of units per afferent subtype. (C) An example microneurography recording of a SA-II unit when gratitude was delivered.

Since only two units were collected for SA-I afferents, we excluded this subtype from the dataset and kept the rest of 39 units. All cutaneous afferents were very sensitive to soft brushing and had mechanical (von Frey) thresholds of activation ≤ 1.6 mN. Per unit, each expression was conducted multiple times, comprising 751 trials in total, with the mean number of trials per unit being 19.26 and the standard deviation being 14.10. Specifically, we collected 128 trials for Field, 151 for HFA, 63 for FA-II, 127 for SA-II, 116 for CT, and 166 for MS. For the six emotion expressions, we collected 135 trials for attention, 124 for calming, 129 for gratitude, 124 for happiness, 119 for love, and 120 for sadness. All recordings were cropped to keep the first 10 s of data (which was the target duration for the trained experimenters) at a resolution of 1 ms. Microneurography data and models are available on Figshare: https://figshare.com/articles/dataset/Models_and_data/257393 10.

## III. DATA ANALYSIS

### A. Afferent Responses to Elementary Touch Gestures

In our first analysis we characterized the firing properties of the afferents in human-delivered touch by comparing the mean instantaneous firing frequency (IFF) and the number of spikes across three elementary touch gestures (tapping, stroking, and holding). The elementary touch gestures were focused here to facilitate comparison with previous studies, summarizing the touch expressions to better align with the contact interactions examined by controlled stimuli, e.g., indentation, brushing, etc. In particular, attention and happiness expressions were grouped as the tapping gesture, calming and love expressions were grouped as the stroking gesture, and the sadness expression was counted as the holding gesture. The gratitude expression was left out since it consisted of both tapping and holding gestures.

Per expression trial, IFF was calculated only at the time point when a spike occurred, defined as the reciprocal of the time duration between the current spike and the previous spike. The mean IFF was derived over the whole 10 s neural recording and the number of spikes was counted from a 1 s chunk containing the largest number of spikes. The duration of 1 s was determined to avoid including long non-contact gaps within the expression. Since multiple emotion expressions and different touch gestures were recorded from the same afferent unit, the Linear Mixed Effects Model (LMEM) was used to perform significance tests on the pairwise comparisons of these variables across afferent subtypes and gestures. Partial $\eta^2$ effect size was calculated for each test and Post-hoc Benjamini-Hochberg method was used for multiple testing correction.

We further examined whether those afferent subtypes can classify touch gestures based on additional aggregated firing features. Five features were extracted as inputs from the 10 s recording of each trial, including the number of spikes, mean IFF, peak IFF, IFF variation, the number of bursts. IFF variation was calculated as the coefficient of variation of IFF, and the number of bursts was defined as the number of spike bursts separated by gaps of inter-spike intervals larger than 1 s. The linear support vector machine (SVM) was implemented for classification with five-fold randomized stratified cross-validation repeated for 20 runs.

### B. Touch Expression Classification

To evaluate the abilities of different afferent subtypes in discriminating the six expressions, we leveraged machine learning classifiers to predict delivered expressions from the neural spike trains. We first developed a one-dimensional convolutional neural network (1D-CNN) for time-series classification. This model was chosen following temporal coding theory [28], which suggests that stimulus information is coded by neurons through the timing and temporal pattern of firing activities. Full 10 s binary spike trains with the resolution of 1 ms were fed as inputs to take the full advantage of the temporal information. The model was trained and tested for each afferent subtype separately. Detailed structure and hyper parameters of the model were determined by cross validation grid search with data from all subtypes combined together. The final model structure contained five CNN layers and 16,646 trainable parameters in total. For each layer, 0.2 dropout was applied. The model was trained based on the loss of categorical cross-entropy with Early Stopping and the ADAM optimizer with a reducing learning rate starting from 0.001. Prediction accuracy was averaged over 20 repeats of five-fold randomly stratified cross validation to obtain more stable results.

To compare with the temporal coding based classification implemented by the CNN, we also employed a linear SVM model for rate coding based classification. Rate coding theory [28] follows that stimulus information is coded by aggregate descriptive features of neural firing, e.g., firing frequency, number of spikes. Therefore, instead of time-series spike trains, the same five firing features used in gesture classification, i.e., the number of spikes, mean IFF, peak IFF, IFF variation, and the number of bursts, were derived from 10 s neural recordings and used as inputs. Moreover, including a simpler model such as SVM is also beneficial since CNN model affords high computational power that might overshadow the encoding capability of each afferent subtype.

### C. Expression Classification with Spike Train Segments

In order to identify the most informative segments of firing patterns that lead to high differentiation accuracies, we further conducted CNN classification on segments of spike trains per afferent subtype. A sliding window method incorporating window position and window length was applied to segment chunks from a given spike train for comparison. The sampling rates for window length were set at intervals of 0.1 s from 0.1 s to 4 s, and at intervals of 0.25 s from 4 s to 10 s, resulted 64 different window lengths in total. For each window length, five segments at different positions were derived according to five metrics with a sliding step of 1 ms, which includes the first segment, the segment with the largest number of spikes, the segment with the highest mean IFF, the segment with the highest IFF variation, and the segment with the highest IFF entropy. The IFF variation and IFF entropy here were calculated from step-interpolated IFF to better reflect the time-series pattern of touch expressions. Therefore, 320 different segment options were obtained in total to be compared.

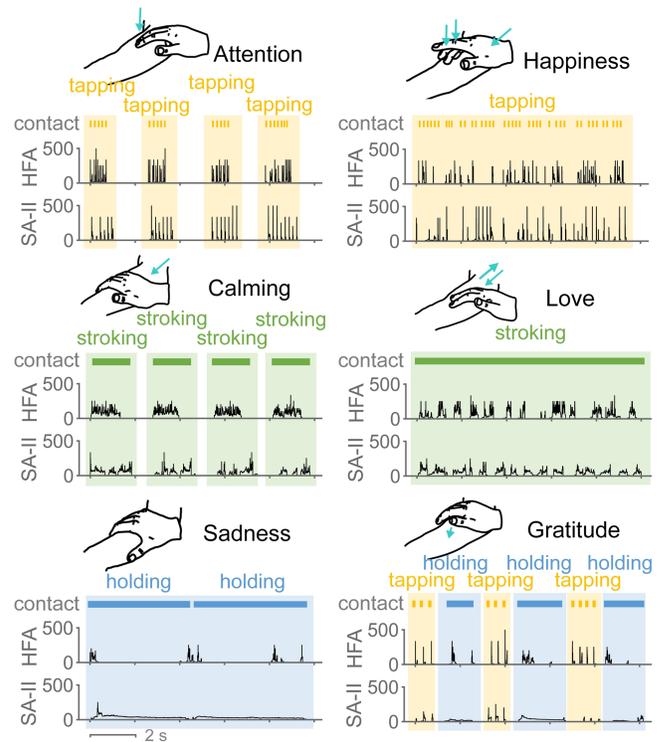

**Fig. 2.** Example microneurography recordings of instantaneous firing frequency (IFF, Hz) collected from HFA and SA-II subtypes when six social touch expressions were delivered on the forearm. Sketches illustrate the standard contact delivery of those expressions. Touch gestures (tapping, stroking, and holding) used by the expressions are denoted.

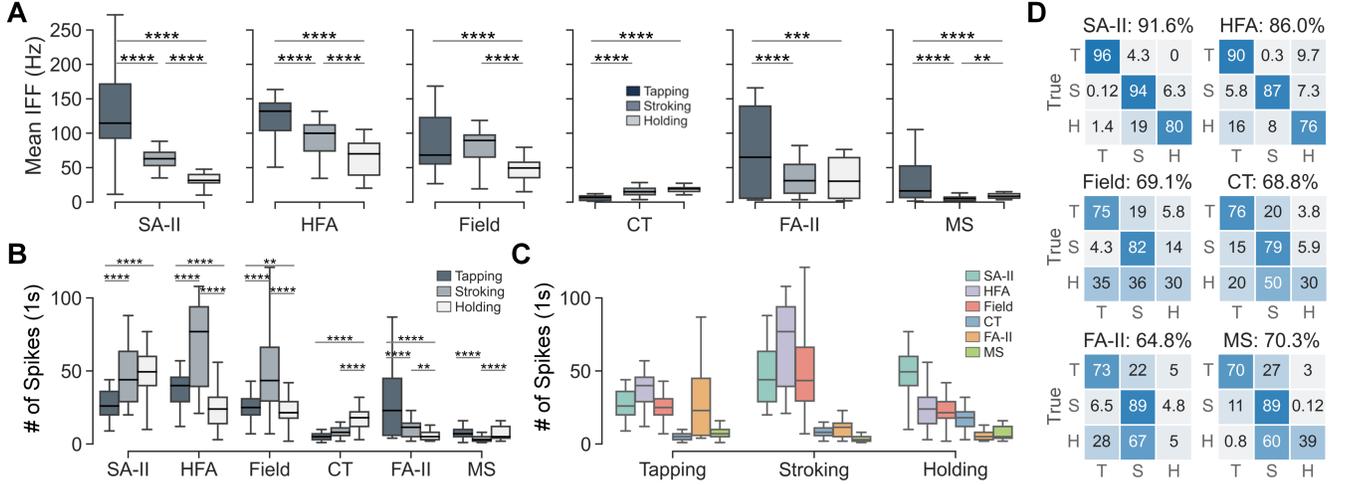

**Fig. 3.** Neural firing properties of the six afferent subtypes in response to different touch gestures. (A) Distributions of mean IFF across gestures per afferent subtype. (B) Distributions of the number of spikes across gestures per afferent subtype. The number of spikes per trial was calculated from a 1 s duration with the largest number of spikes. (C) Distributions of the number of spikes across afferent subtypes per gesture. *p < 0.05, **p < 0.01, ***p < 0.001, ****p < 0.0001. Significance test results for panel C and partial $\eta^2$ effect sizes for panels A, B, and, C are provided in Supplementary Fig. S1. (D) SVM classification of touch gestures per afferent subtype using five firing features extracted from 10 s recordings. "T" represents tapping, "S" represents stroking, and "H" represents holding. Numbers in the cells denote the percentage of classification results.

For each afferent subtype, we first investigated the best window position metrics by conducting CNN with five-fold randomly stratified cross-validation repeated twice. Prediction accuracies of the five metrics were averaged across all window lengths, where Mann-Whitney U tests and post-hoc Benjamini-Hochberg correction were applied for pairwise comparison. Based on the best two window position metrics per subtype, we examined the prediction accuracies along with the window length by conducting CNN with seven repeats of five-fold randomly stratified cross-validation. We identified the saturation window length per subtype based on 90% of the highest accuracy from its accuracy curve fitted by fourth-order polynomial regression. Accuracy curves and saturation window lengths were further derived for all expressions per afferent.

According to the best window position and the saturation window length, the most informative firing segments were identified per afferent (group 1) and per afferent-expression combination (group 2). SVM classification was then conducted on the identified segments to examine if they also yield high prediction accuracies when using only five aggregated firing features and a linear model.

### D. Expression Classification with Spike-Timing Noise

We aimed to further evaluate the contribution of the fine-grained temporal information present in the spike train to the accuracy of time-series classification. We therefore examined the spike-timing sensitivity of all afferent subtypes in classifying touch expressions. Random noise was added to all spike times across the full 10 s spike trains, which were then input to the CNN classifier. Noise following a Gaussian distribution was employed with mean equal to zero and standard deviation (SD) ranges from 0 to 100 ms with steps of 5 ms. The CNN model was trained per subtype with noise-free spike trains and was tested using recordings with noise added. Average accuracies were obtained per combination of afferent subtype and expression from five repeats of five-fold randomly stratified cross validation, with each level of noise tested by ten different sets of random noise.

### IV. RESULTS

### A. Firing Properties of Afferent Subtypes

Examples of collected neural recordings of SA-II and HFA afferents are illustrated in Fig. 2 for all six touch expressions. Despite the consistent delivery of the expressions, distinct firing patterns were observed between these two subtypes. For example, with the sadness and gratitude expressions, SA-II afferents responded throughout contact with a sustained, slowly decaying firing pattern, while HFA afferents only responded to the onset and offset of the holding or when the hand position was adjusted.

Under human-delivered social touch, all afferent subtypes exhibited similar ranges of mean IFF and number of spikes (Fig. 3) compared with the same subtypes recorded with controlled stimuli [8], [16], [19], [33]–[36]. More specifically, the mean IFFs of Aβ afferents (up to 300 Hz) were overall higher than those for CT and MS afferents (up to 50 Hz) (Fig. 3A), similar to prior studies using passive touch [8], [33], [36]. Also, the mean IFFs of SA-II, HFA, and Field afferents decreased when switching from fast tapping to stroking to static holding contact, yet increased for CT afferents (Fig. 3A). For stroking contact alone, it has been reported that the mean IFFs of Aβ afferents increased with higher velocity, while the mean IFFs for CT afferents decreased for velocities over 3 cm/s [8]. As for the number of spikes, HFA and Field afferents shared the same patterns, with stroking contact eliciting significantly more spikes and holding contact eliciting fewer spikes (Fig. 3B). Note that fewer spikes recorded from tapping contact may be due to the overall shorter contact duration relative to the other two gestures. In comparison, the numbers of spikes for SA-II and CT afferents were also high for slow and static holding contact, which agrees with the firing properties widely reported for these two subtypes [8], [16], [34]. Overall, such alignments in firing properties compared with those identified using controlled stimuli help validate the effectiveness of the designed microneurography paradigm and experimental procedure of human social touch.

Meanwhile, all Aβ afferents subtypes in the skin responded very well to tapping contact (Fig. 3C, p-values in Supplementary Fig. S1), while SA-II responded with significantly more spikes for holding than other gestures and FA-II exhibited significantly fewer spikes for stroking than other gestures. These distinct properties suggest the potential complementary functional roles of those afferents when viewed as a population at higher levels of the nervous system. Moreover, when five aggregated firing features were used (see section III-A), the three elementary touch gestures can be well classified by all afferent subtypes (Fig. 3D) with the highest accuracies obtained by SA-II and HFA afferents.

## B. Single Units of SA-II and HFA Afferents Effectively Classify Social Touch Expressions

Among the six afferent subtypes, SA-II and HFA achieved the highest accuracies around 70-80% in classifying the six touch expressions using the CNN with full spike trains as inputs (Fig. 4A). Note that the results may slightly vary across different runs due to the random train-test splitting and stochasticity of CNN model. Such accuracies are very close or even slightly higher than human recognition accuracy for the same six standard touch expressions [4]. In comparison, Field afferents exhibited relatively lower accuracy around 56%, while the accuracy of CT, PC, and MS afferents were not far from the chance level of 16.7%. Similar prediction results were also obtained in SVM classification when using five firing features (Fig. 4B). Classification accuracies as high as 70-80% were observed for SA-II and HFA subtypes while Field, CT, FA-II, and MS afferents exhibited lower accuracies. The consistency in classification performance between the two models implies that SA-II and HFA afferents convey the richest information in human social touch for at least the tested six touch expressions and are capable of encoding the mechanical skin deformations relevant to social touch expressions in an accurate and reliable way.

## C. Most Informative Firing Patterns

Among the five window position metrics used in generating spike train segments, significantly different classification accuracies were observed among most pairs of metrics across all subtypes (Fig. 5B, p-values in Supplementary Fig. S2B). For comparison purposes, we picked the best two window position metrics per subtype: the highest number of spikes and the highest mean IFF for SA-II, first and the highest number of spikes for HFA, and the highest number of spikes and the highest IFF entropy for Field, CT, FA-II, and MS. The accuracy differences among the window metrics were relatively small, such as 3.3% between the top two metrics for SA-II afferents, and 2.7% between the top two metrics for HFA afferents. Moreover, five accuracy curves along with the window lengths corresponding to the five window metrics also well overlapped with each other, especially for SA-II and HFA subtypes (Supplementary Fig. S2C). It indicates that window position does not make a huge impact on the classification performance.

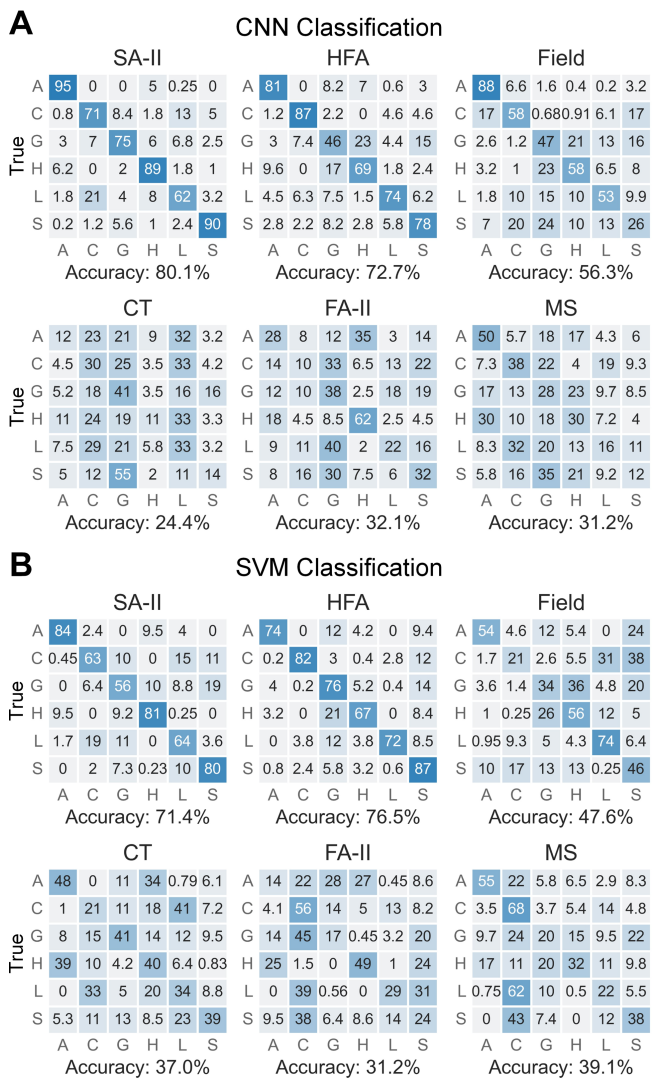

**Fig. 4.** Classification results per afferent subtype using (A) CNN classifier with 10 s spike trains as inputs and (B) SVM classifier with five features as inputs. Numbers in the cells denote the percentage of classification results. The classification accuracy is markedly higher for SA-II and HFA subtypes, at levels observed in human perceptual experiments [4].

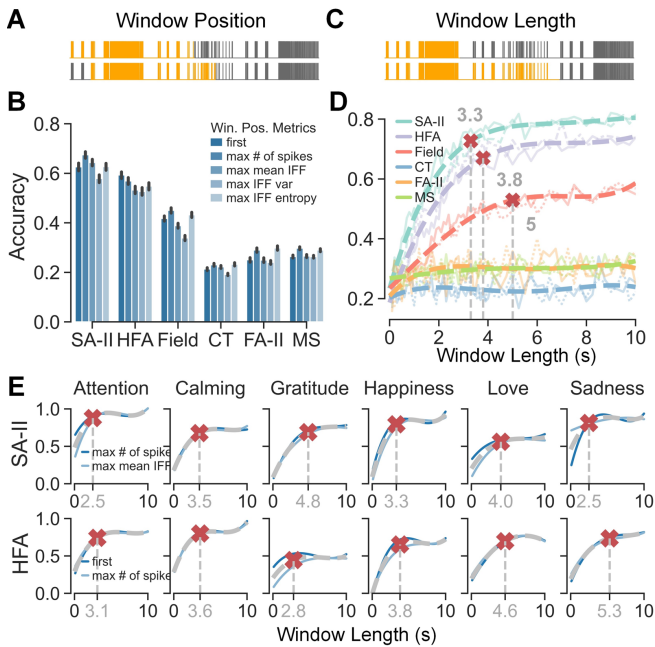

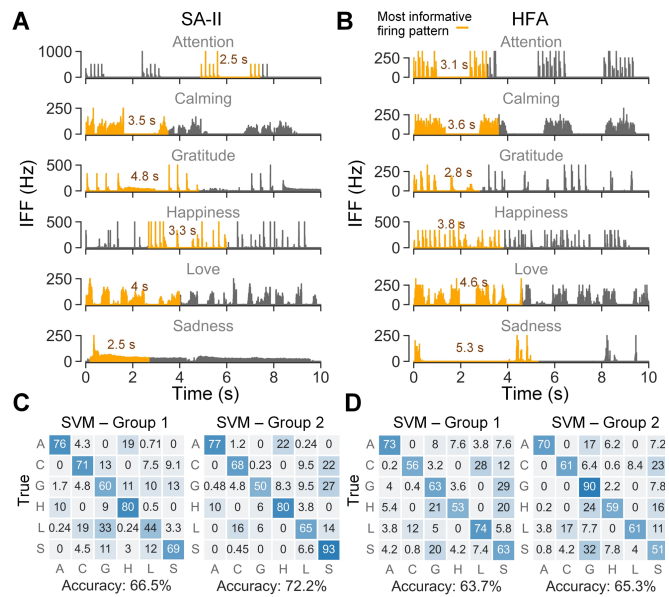

**Fig. 5.** Comparison of CNN classification accuracies when using segments of spike trains derived from different window positions and window lengths. (A) An example of two window position options with the same window length. Gray traces are 10 s spike trains from the same trial, where highlighted spikes illustrate two different segments. (B) Classification accuracies across window position metrics averaged over all window lengths for each afferent subtype. Significance test results can be found in Supplementary Fig. S2B. (C) An example of two window length options with the same window position. Gray traces are 10 s spike trains from the same trial, where highlighted parts illustrate two different segments. (D) Classification accuracies along with window length per afferent subtype. Accuracy curves were fitted using data from their best two window positions and fourth-order polynomial regression, shown as dashed curves. Red cross markers denote 90% saturation window lengths. Two lighter curves represent data from the two best two window positions. (E) Classification accuracies along with window length per afferent subtype per expression. Averaged accuracies from their best two window positions are shown as grey dashed curves and blue curves represent each of the best positions. Red cross markers denote 90% saturation window lengths.

Per afferent subtype, the two top performing metrics were adopted for examining the influence of window length. Results show that classification accuracies for SA-II, HFA, and Field afferents saturate when window length approached 3.3 s, 3.8 s, and 5 s respectively (Fig. 5D). In contrast, accuracies for the other afferent subtypes began and remained consistently low. It implies that instead of the full 10 s, an average duration of 3-4 s of the neural responses of SA-II and HFA afferents provides sufficient information to differentiate those expressions. Further inspection into afferent-expression combinations shows that saturation window lengths varied between 2.5 s to 5.3 s across expressions for SA-II and HFA afferents, which is still a comparably limited range much less than 10 s.

The identified most informative firing patterns of SA-II and HFA afferents are shown in Figs. 6A, 6B for all expressions. We found that for expressions with multiple rounds of contact, e.g., attention and calming, at least one round of contact was always included, which captures the specific rhythm of the contact delivery across different expressions. Moreover, the variation of the saturation window length could be related to both contact stimuli and firing properties of afferent subtypes. For example, the unique repetitive tapping pattern of attention expression might explain why it requires relatively less data than other expressions. Sadness exhibits the largest difference in saturation window length between SA-II and HFA (Fig. 5E). One explanation is the sustained low-frequency firing pattern of SA-II afferents under holding contact is easy to differentiate even within a shorter time. In comparison, the firing response of HFA to holding is similar to that of tapping contact such that more data including the non-response gap are needed to capture the unique dynamic of prolonged holding of the sadness expression (Fig. 6B).

**Fig. 6.** Examples of identified most informative firing patterns and SVM classification based on the identified segments. IFF traces highlighted in orange are the most informative segments determined by the best window position (SAII: max # of spikes, HFA: first, derived from Fig. 5B) and the saturation window length per afferent-expression combination (annotated as numbers near the highlighted segments, derived from Fig. 5E) for (A) SA-II subtype and (B) HFA subtype. Classification results using the SVM model for (C) SA-II subtype and (D) HFA subtype. Group 1 and group 2 refer to two groups of spike train segments derived by saturation window lengths per afferent subtype (Fig. 5D) and saturation window lengths per afferent-expression combination (Fig. 5E), respectively.

Moreover, SVM classification using the identified most informative firing patterns show similar prediction accuracies as using the full 10 s recordings (Fig. 6C, 6D). Slightly higher accuracies were obtained by segments derived per afferent-expression combination (group 2). Such findings help validate the richness of information contained within identified segments of SA-II and HFA afferents' firing patterns.

### D. Spike-Timing Sensitivity

We found that SA-II afferents were sensitive to spike-timing noise for attention, happiness, sadness, and gratitude expressions as their prediction accuracies dropped to approximately a chance level when noise higher than 50 ms SD was applied (Fig. 7B). Those four expressions were delivered by tapping and holding gestures, while expressions delivered by the stroking gesture, i.e., calming and love, were not influenced by spike-timing noise. HFA afferents were sensitive to spike-timing noise for only tapping-delivered expressions of attention and happiness. For other touch expressions delivered by holding or stroking gestures, their prediction accuracies did not drop when noise increased. Compared with HFA afferents, SA-II's unique spike-timing sensitivity to holding contact indeed align well with its unique sustained response to static contact. Except for SA-II and HFA afferents, other afferent subtypes were not sensitive to spike-timing noise across all expressions. Moreover, SA-II and HFA subtypes also exhibited tolerance to a lower level of spike-timing noise. More specifically, SA-II responses to tapping contact were tolerant to spike-timing noise up to 20 ms. In comparison, responses to holding contact began to be influenced at roughly 10 ms. For HFA afferents, responses to tapping contact exhibited noise tolerance up to approximately 15 ms. This tolerance could relate to the variability of human-delivered touch, the variability of firing patterns across units, and/or the prediction target of expressions being abstract and composite.

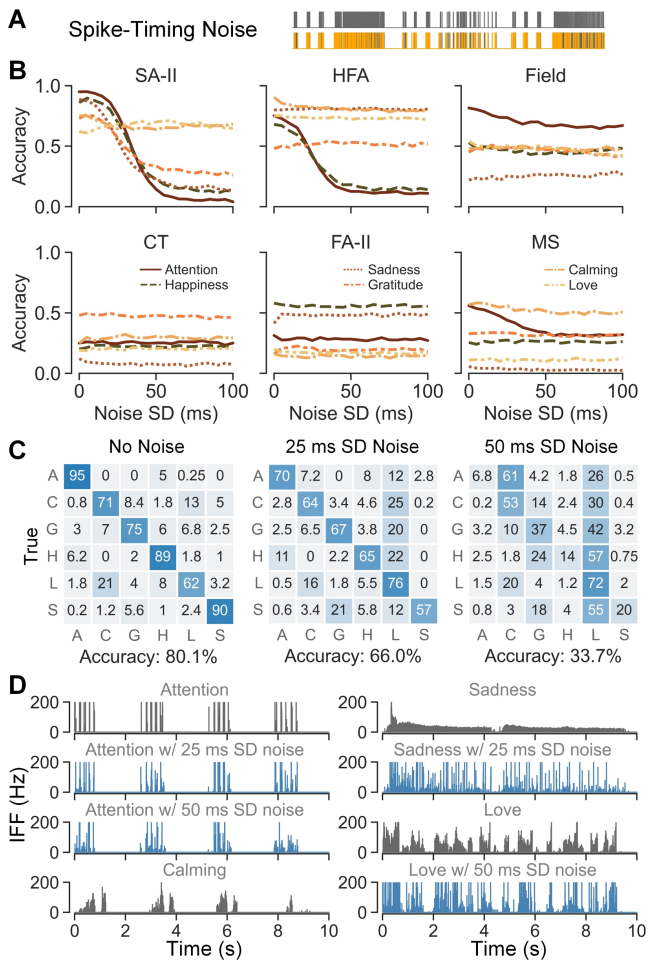

**Fig. 7.** Spike-timing sensitivities across afferent subtypes in human social touch. (A) Spike trains from the same trial with (lower) and without (upper) spike-timing noise added. (B) CNN classification accuracies of six expressions relative to the standard deviation of added noise per afferent type. (C) CNN classification accuracies for the SA-II subtype using 10 s spike trains with different level of noise added. (D) SA-II recordings for the confused expressions when noise is added. The grey IFF traces are the original recordings and the blue IFF traces are recordings with noise added.

We then focused on SA-II afferents to investigate the potential cause of such high spike-timing sensitivity of certain afferent-expression combinations. Confusion matrices derived from CNN classification (Fig. 7C) show that the tapping contact with the attention expression was misclassified as stroking contact of the calming expression when 50 ms SD noise was applied, while holding contact of the sadness expression was misclassified as stroking contact of love when 25 ms noise was applied. Neural recordings with and without noise were next compared for those two confused cases (Fig. 7D). We found that, for attention expression, noise as high as 50 ms SD could flatten isolated spikes elicited by repetitive taps within one round of tapping. It thus changed the pattern envelope to a continuous chunk of firing with variable frequencies, which was similar to the firing pattern of calming delivered by stroking. As for the sadness expression, 25 ms SD noise already converted its sustained slowly decaying firing pattern into a spiky and irregular shape, which was similar to the firing pattern of love delivered by stroking. Here, attention and calming were mainly confused with the stroking of calming and love respectively, which could relate to their shared touch rhythm of having prolonged non-contact gaps or not. In contrast, since SA-II responses elicited by stroking contact were initially irregular, noise as high as 50 ms SD still did not cause a major change to the firing response of the love expression. Based on the observations, we hypothesize that the spike-timing sensitivity of those afferent subtypes could be strongly tied to the extent of changes in the envelope of their firing patterns caused by noise. This pattern envelope could be a more appropriate metric in capturing contact pattern at a macro level, such as touch gestures and contact rhythms when encoding touch expressions. In this scenario, millisecond-precision of single spike times might not be as informative due to robustness of the touch expressions and their social meanings.

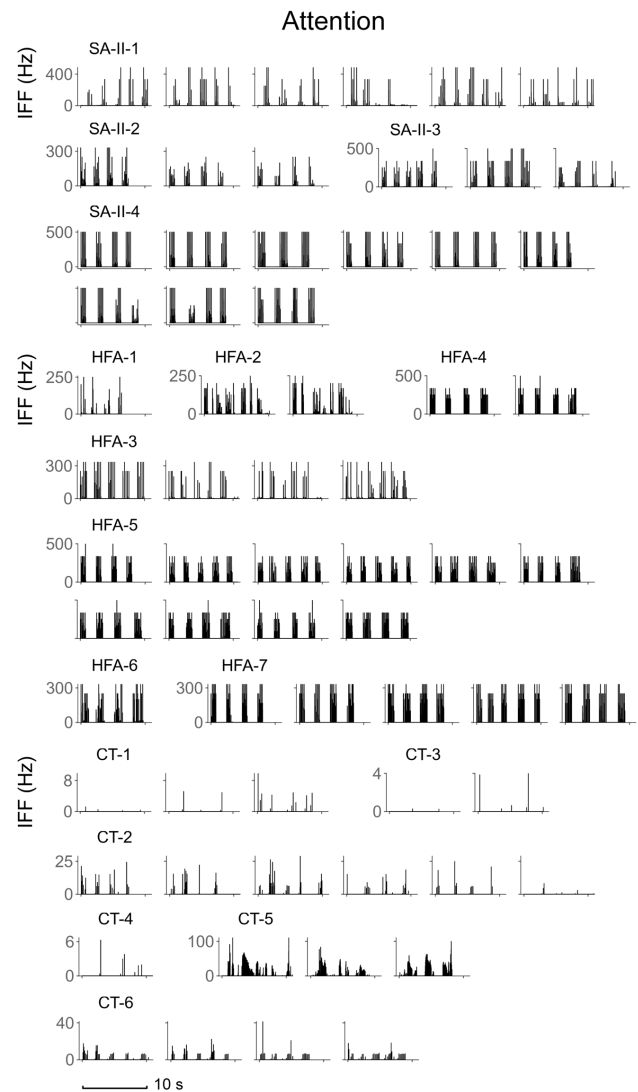

**Fig. 8.** Neural recordings of SA-II, HFA, CT subtypes for the attention expression. All trials are shown here in the format of Instantaneous firing frequency (IFF, Hz).

## V. DISCUSSION

### A. Microneurography Paradigm for Human-to-Human Touch

Distinct from traditional experiments that control the mechanical stimulus and vary a single feature at a time, we record from single peripheral afferents in a human-to-human touch paradigm, where multiple stimulus features, e.g., normal displacement, contact area, lateral velocity, vary simultaneously [1], [4], [11], [37]. Such naturalistic human touch interactions directly contribute to our emotional well-being and maintains our social connections [38], yet are technically difficult to replicate with actuated devices. Indeed, precisely controlled stimuli, such as rigid bodies indented in one dimension of depth or force [17]–[19], are more commonly employed in characterizing the firing properties of peripheral afferents. Recent efforts have begun to move toward more naturalistic contact interactions using brushing, puffs of air, and pinch, etc., [39], [40]. Natural textures have also been applied in recording monkey Aβ afferents [26]. However, each of these efforts still controls and varies a single stimulus feature at once, which is different from natural contact with co-varying features.

In this study, we move a step further into human-delivered touch, where the richness of contact dynamics could reveal classes of primary afferents that encode the combination of multiple features. In our tasks, such information could be relevant to social messages conveyed in touch expressions. More specifically, six standardized social touch expressions were delivered by trained experimenters. This affords reliable contact interactions [4] and retains the subtleness of human-delivered touch at the same time. Meanwhile, expressions were designed with specific touch gestures, which can be compared with similar mechanical stimulus contact, e.g., human-delivered stroking versus brush-delivered stroking, human-delivered tapping versus vibrating actuator indentation. Indeed, the firing properties we observed in human touch (Fig. 3) share similar ranges and trends with those for controlled stimuli. It also demonstrates that similar states of skin contact and deformation could elicit similar responses across human touch and stimulus contact [12].

### B. Social Touch Relevant Encoding across Afferent Subtypes

Both CNN classification using time-series neural recordings and SVM classification using five firing features show that SA-II and HFA subtypes outperform other subtypes (Fig. 4) and provide high differentiation accuracies similar to human perception [4]. Moreover, such accuracy is consistent in using either the full 10 s time course of the neural responses or the most informative firing patterns therein (Fig. 6C, 6D).

The SA-II and HFA afferent subtypes, due to particular physiological mechanisms, may be geared more to the inherent contact characteristics of social touch. One prominent commonality is their large, but not too diffuse, receptive fields [30], [41], [42], which may help in consistently capturing the range of contact dynamics given the size of touchers' fingers and hands and their lateral movements. The detailed sizes of receptive fields of those mechanoreceptive Aβ afferents have been reported by a series of microneurography studies [30], [41], [42]. In particular, Vallbo et al., [30] has recorded relatively large receptive fields for rapidly adapting units in the hairy skin of human forearm, with around 113 $mm^2$ for HFA and 78 $mm^2$ for Field afferents. On the hairy skin of human hands [42], median sizes were identified as 16 $mm^2$ and 28 $mm^2$ for SA-I and SA-II units respectively. In glabrous skin [41], the receptive field sizes of SA-II afferents have also been shown to increase considerably with indentation force, as compared with SA-I units. FA-II afferents that innervate Pacinian corpuscles, on the other hand, exhibited markedly larger receptive fields [42], which are almost too diffuse to map due to their extreme sensitivity [19]. Therefore, compared with other subtypes, the relative size of the receptive fields of HFA and SA-II afferents in hairy skin could contribute to their social expression encoding.

Furthermore, SA-II and HFA afferent subtypes are believed to be sensitive to a wide range of contact, including indentation [43], hair deflection [44], skin stretch [42], and shearing forces [16], [45], which are contact characteristics that human touch gestures tend to evoke. For example, both SA-II and HFA respond to tapping (vertical contact) and stroking (sheering contact) with distinct mean IFFs (Fig. 3A) and can easily differentiate those two gestures (Fig. 3D). In contrast, Field and FA-II afferents respond to these two directions of contact with non-differentiable firing frequencies (Fig. 3A). Moreover, SA-II and HFA afferents also precisely followed tapping contact with high IFF responses (Fig. 8), outperforming the other subtypes. Interestingly here, SA-II afferents are typically thought to mainly encode static/slow movements and skin stretch [16], [45], but also responded very well to fast vertical contact delivered by human tapping. As for holding contact, as expected, SA-II afferents respond with sustained low-frequency firing patterns, which distinguish holding from other fast movements. HFA afferents did not respond to the sustained contact, but precisely captured the on-set and off-set of the hold gesture. Although this pattern of spike firing is similar to that of tapping, the unique prolonged touch rhythm of holding provides distinct temporal information. Meanwhile, those two subtypes also respond to the stationary holding gesture with a significantly lower mean IFF compared with other gestures. Overall, the capability of SA-II and HFA subtypes to differentiate the social touch expressions suggests that their neural responses well correspond to the range of stimulus input and mechanical skin deformation inherent in human-to-human touch interactions.

Focusing on the context of social touch, the afferent subtypes exhibited distinct sensitivities in encoding the two layers of information, i.e., gestures (lower level) and expressions (higher level). Based on the same five firing properties, all six subtypes could accurately differentiate the three gestures (Fig. 3D), whereas CT, FA-II, and MS afferents fail to separate the expressions (Fig. 4). It suggests that contact patterns of elementary touch gestures, e.g., tapping, stroking, and holding, can be captured to a certain extent by all afferent subtypes. While the same gesture can be slightly varied in terms of its contact delivery of velocity, indentation depth, contact area, etc. [11] to convey specific social meanings, such nuances may be less easy to capture for certain afferents. For example, attention and happiness delivered by tapping, and calming and love delivered by stroking, were frequently confused by CT, FA-II, and MS subtypes as they share comparable contact dynamics. This might also explain human receivers' misidentification of those expression pairs [4]. In comparison, SA-II and HFA subtypes are very sensitive to slight contact changes, as they classify gestures and expressions with relatively similar accuracies (Fig. 3D, 4).

The relatively lower coding capability of CT, FA-II, and MS afferents might relate to their functional roles in signaling contact modalities less reflected in the applied six social touch expressions. CT afferents are traditionally thought to signal affective touch, more specifically pleasantness elicited by slow and gentle stroking [6], [46], in parallel with Aβ afferents serving as discriminators for physical contact properties [47]. We indeed found that CT afferents can successfully identify stroking contact (Fig. 3D), yet could not further differentiate contact between love and calming expressions (Fig. 4). More specifically, gentle stroking was deployed for both expressions but with different contact rhythms and routes, i.e., love: continuous back-forth stroking, calming: four separate one-direction stroking. Meanwhile, although CT afferents have been recorded with relatively small receptive fields [43], Olausson et al., [48] reported that neuronopathic patients lacking Aβ afferents exhibited a poor ability to localize tactile stimuli based on CT afferents alone. Such weak contribution of CT afferents to localization perception, along with their low sensitivity to very fast movements [8], suggests that the combination of Aβ afferents might be needed to inform subtle contact differences. Surprisingly, CT afferents also respond very well to fast vertical tapping contact (Fig. 8). While CT afferents have been reported to respond well to von Frey indentation [49], human tapping affords much higher levels of force in a faster and repetitious manner. However, more detailed contact differences between the tapping of attention and happiness were not captured. For the other two subtypes, FA-II afferents respond to high-frequency vibration in discriminative touch, such as contact delivered to a site remote to the afferent's receptive field center [42]. However, they filter low frequency stimuli [19] that carry most of the information adhering to social touch. MS afferents respond to muscle extension and flexion associated with our proprioceptive sensation [50] and while they could discriminate the lower level gestures (Fig. 3D), their relatively low firing rates (Fig. 3A-C) suggest that our social touch stimuli are not optimal stimuli.

## C. Temporal Envelope of Firing Pattern as Potential Social Touch Encoding Strategy

By leveraging machine learning classification models, we identified the most informative firing patterns of SA-II and HFA afferents in encoding touch expressions. Those firing patterns and their corresponding contact patterns suggest their coding strategies relate to perceptual discrimination. More specifically, instead of the full 10 s of contact, we found that an average of 3-4 s provides enough information for single units to differentiate the six expressions (Fig. 5D). Also, as window position did not have a critical impact, it suggests that afferents respond in a consistently informative way throughout the course of contact, where the accumulation of a sufficient amount of information would be the key for social touch processing. Indeed, this time duration of 3-4 s aligns with the cortical response time of brush-delivered affective touch [51], facial EMG response time in natural social touch that reflects emotional processing [4], and the acceptable response time of humanoid robots being touched by a human [52]. However, this time duration is significantly longer than that reported in encoding precisely-controlled single stimulus features. Based on a population simulation of peripheral tactile afferents, tens of milliseconds were found to be sufficient in encoding stimulus directions [53]. Similarly, with ensembles of recorded single afferents, tens of milliseconds neural responses were also suggested to be effective in encoding controlled force, torque, force direction, and shape on finger pads [54], [55]. Such a difference in time course highlights the complexity of human social touch, where social meanings may be integrated from specific spatiotemporal contact interactions. For example, attention was expressed as separate rounds of repeated fast tapping versus happiness was expressed as continuous tapping with multiple fingers tickling back and forth on the forearm. While afferent firing at tens of milliseconds can be comparable between the two expressions, they begin to reflect the contact rhythms of expressions in the time scale of multiple seconds (Fig. 6A, 6B). In comparison, controlled single stimulus features carry less information and thus could be identified with much shorter neural responses especially using a population model [53]. As single units were tested in our case, we expect that population responses of single or multiple afferent subtypes might encode social touch expressions in shorter durations.

Furthermore, 10 to 20 ms SD random shifts applied to the spike timing cause little effect on the classification of the expressions, although greater shifts can change the firing envelope of one expression to be confused with another. It appears that the spike timing precision needed in encoding human social touch is relatively lower than encoding controlled stimulus features. For instance, when classifying well-controlled scanned textures and vibratory stimuli, the optimal spike timing precision is around 1 to 10 ms [26], [56]. Although the distance of transforming one spike train exactly to another [57] was used in aforementioned studies, we directly added artificial jitter to spike times [58]–[60] given the variation of human contact delivery. It was believed that spike-timing jitter would blur the transmitted information of the stimulus [58], [59], [61]. For encoding controlled audio amplitudes, milliseconds or even sub-milliseconds of added artificial jitter can significantly decrease the accuracy of transmitted information [60]. Therefore, in human social touch, the relatively higher tolerance to spike-timing jitter suggests that the coarser level of temporal pattern might play a role. We found that by adding spike timing jitter, the envelope of the firing pattern can be drastically changed, which may be closely related to macro level touch interaction of gestures (Fig. 7D), instead of cell level dynamics of signal transmission. It also aligns with the finding that the SVM model using aggregated firing features provided comparable classification performance as the time-series CNN model (Fig. 4). It implies that detailed spike-to-spike temporal coding may not contribute to the core information in human social touch scenarios. Meanwhile, rate coding of aggregated features might not capture the whole dynamic details. Here we hypothesize that the temporal envelope of the firing pattern, which falls between the precise temporal coding and the summarized rate coding, could be a valuable metric in encoding social touch expressions, where the window length would have a large impact. It also offers references for future designs of social touch devices, including potential durations and resolutions for haptic rendering.

### D. Limitations and Future Works

The slowly-adapting type I (SA-I) afferent is another Aβ subtype that is likely to play a significant role in encoding social touch stimuli. In general, SA-I afferents contribute to our abilities in fine touch discrimination [62]. In our study, the population of SA-I afferents (n=2) was not large enough to include. Our speculation is that SA-I afferents might behave akin to SA-II afferents, due to similar adaptation characteristics. Additionally, SA-I afferents exhibit a very large dynamic range of sensitivity, as compared to SA-II afferents, combined with very low absolute thresholds [63]. Such sensitivity should benefit discriminability in general, yet if SA-I afferents are too sensitive, this may be too variable a response than buries the core contact information carried in social touch. In this way, SA-II afferents might offer advantages because they have relatively dampened responses to dynamic stimuli compared to SA-I afferents. Indeed, less sensitive subtypes in high threshold mechanoreceptors are shown to better encode noxious forces than the more sensitive ones [23]. However, further follow up work is required to understand the response characteristics of the SA-I subtype to social touches.

Since peripheral afferents convey all mechanical sensory inputs to the nervous system, one perspective is that the expressions simply reflect arbitrary collections of different mechanical inputs. However, social meanings assigned to the six expressions together with the discriminability differences across those subtypes can be interpreted as the evidence for early tuning of the nervous system to facilitate interpretation of social touch. Our study concerns to what extent the first stages of the nervous system would provide the scaffolding for the complex neural processing of social touch. We have shown here that at least two afferent subtypes, SA-II and HFA, provide more information than others. As different subtypes are believed to be responsible for certain mechanical features, our results imply that mechanical tuning properties of those two subtypes are particularly well suited for the contact dynamics embedded in social touch expressions. We predict from our results that these two subtypes would have a stronger influence than other classes in the neural pathway of social touch. Such findings also provide insights into haptic rendering of social touch, where contact stimuli preferential for SA-II and HFA afferents could be prioritized. Meanwhile, while single units appear to hold discriminative capacity, afferent subtypes are likely to interplay in a cohesive way in generating population responses [27], [64], from which our perception and discrimination are gleaned. Our findings regarding single unit responses provide the foundation for such future explorations, where empirical or mathematical studies of higher-order nervous structures would be needed to unravel the population processing of social touch communication.

Additionally, at the single-unit level, it is possible that SA-II and HFA afferents may struggle to distinguish different sets of touch expressions than those we used, and other subtypes may excel. However, it is empirically challenging to include a large set of emotional messages in human touch microneurography experiments while maintaining the data size for effective analysis. The six emotional messages adopted here were reported to be easily recognizable through touch [1], [3], [4], [65], [66], while many others are difficult to communicate using touch alone. In our study, only one expression was used per emotional message and was constrained to be delivered on the forearm. The forearm was chosen for the benefits of microneurography setup as well as for being the body portion widely acceptable and studied in social touch scenarios [67]–[71]. The specific expression was derived from the commonly adopted touch behaviors of that emotion that is understandable by human receivers [1], [4]. Among the selected six touch expressions, a wide range of contact dynamics were included with varying velocities, movement directions, contact areas, indentation depths, etc., [4], [11]. That said, it may be beneficial to vary expressions per emotion in future studies, which could also take into account individual differences in touch delivery [13] and emotion perception [4].

Meanwhile, with the expressions connected to specific social meanings, the underlying emotional contexts could be moderated. In particular, the perception of pleasantness (valence), emotional arousal, and dominance [72], [73] are not fully explored in this study. Part of the reasons was to avoid the high task load of participants if psychophysical and microneurography experiments were conducted together. Based on the dataset of emotional ratings for English words [73], happiness and attention afford high arousal and were found both delivered by fast tapping contact. We might assume that neural responses to fast contact velocities are related to high arousal percepts. However, other contact characteristics, e.g., force, indentation, contact area might also contribute [11]. Therefore, precise contact quantification needs to be introduced to uncover further details of how emotional contexts of physical touch delivery are encoded by peripheral afferents [37].

## VI. CONCLUSION

In this work, through microneurography recordings of single peripheral afferents elicited by naturalistic, human-delivered social touch, we found that Aβ afferents, especially SA-II and HFA subtypes, can effectively encode social touch expressions. Indeed, the responses of single afferents match the discriminative accuracy of human perceptual recognition. More specifically, the analysis of spike firing patterns using time-series machine learning classification indicates that a duration of 3-4 s of spike firing provides sufficient discriminatory information in social touch, with high tolerance to shifts in spike-timing of 10-20 ms, suggesting the time scales relevant for the peripheral encoding of social touch interactions are distinct from millisecond accuracy requisite in discriminative touch interactions.

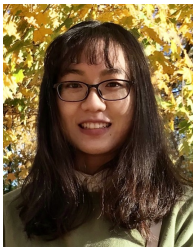
Shan Xu received her BE degree in Intelligent Science and Technology and her ME degree in Control Science and Engineering from Nankai University, China, in 2016 and 2019, respectively. She is currently working toward her PhD degree in Systems Engineering at the University of Virginia.

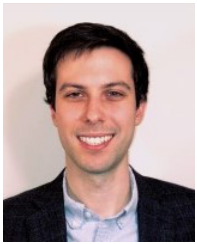
Steven Hauser received his PhD degree in Systems Engineering from the University of Virginia in 2019. He also received his MS degree in Biomedical Engineering and his BA degree in Computer Science from the University of Virginia, in 2016 and 2014, respectively. He is currently a senior data scientist with Size Stream.

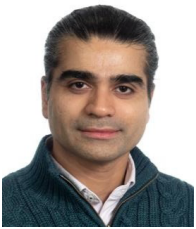
Saad Nagi received his BMedSc(Hons) and PhD degrees from the School of Medicine of Western Sydney University. He is currently an associate professor with the Department of Biomedical and Clinical Sciences and the Center for Social and Affective Neuroscience, Linköping University, Sweden.

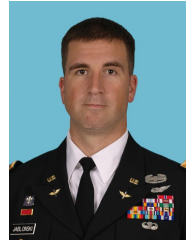
James Jablonski received his Bachelor's degree in Physics from the United States Military Academy at West Point, his Master's degree in Operation Research from the U.S. Air Force Institute of Technology, and his PhD degree in Systems Engineering from the University of Virginia. He is currently a director with TRAV Monterey.

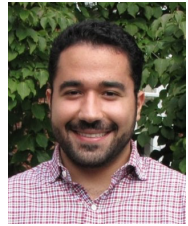
Merat Rezaei received his BS in Mechanical Engineering from Idaho State University, in 2018, and his MS in Systems Engineering from the University of Virginia, in 2021. He is currently working toward his PhD degree in Systems Engineering at the University of Virginia.

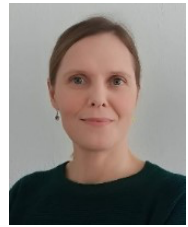
Ewa Jarocka is currently a principal research engineer with the Department of Biomedical and Clinical Sciences and the Center for Social and Affective Neuroscience, Linköping University, Sweden.

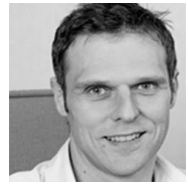
Andrew Marshall is currently a senior lecturer in Pain Neuroscience and an honorary consultant clinical neurophysiologist with the University of Liverpool, United Kingdom.

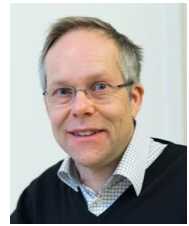
Håkan Olausson is a full professor in Neuroscience with Linköping University. He is the principal investigator of the Department of Biomedical and Clinical Sciences and the Center for Social and Affective Neuroscience, Linköping University.

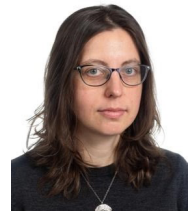
Sarah McIntyre is an Assistant Professor with the Department of Biomedical and Clinical Sciences and the Center for Social and Affective Neuroscience, Linköping University, Sweden. She received her BS in Psychology and her PhD from the University of Sydney.

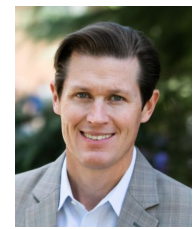
Gregory Gerling is a full professor in Systems Engineering with the University of Virginia. He received his BS in Computer Science from the University of Iowa, his MS in Industrial Engineering and his PhD in Industrial Engineering from the University of Iowa.

## A

### Partial η² Effect Sizes for Fig. 3A, 3B

| Mean IFF | SA-II | HFA | Field | CT | FA-II | MS |
|---|---|---|---|---|---|---|
| Tapping - Stroking | 0.52 | 0.53 | 0.00059 | 0.24 | 0.33 | 0.23 |
| Stroking - Holding | 0.66 | 0.8 | 0.36 | 0.52 | 0.32 | 0.23 |
| Tapping - Holding | 0.72 | 0.47 | 0.56 | 0.048 | 6.1E-5 | 0.12 |

| # of Spikes | SA-II | HFA | Field | CT | FA-II | MS |
|---|---|---|---|---|---|---|
| Tapping - Stroking | 0.51 | 0.68 | 0.4 | 1.1E-5 | 0.32 | 0.32 |
| Stroking - Holding | 0.6 | 0.27 | 0.078 | 0.32 | 0.42 | 0.0022 |
| Tapping - Holding | 3.9E-5 | 0.75 | 0.39 | 0.36 | 0.18 | 0.38 |

## B

### LMEM Test Results and Partial η² Effect Sizes for Fig. 3C

| Tapping | SA-II | HFA | Field | CT | FA-II | MS |
|---|---|---|---|---|---|---|
| SA-II | | 0.12 | 0.026 | *<br>0.37 | 8.8 E-5 | ****<br>0.7 |
| HFA | | | 0.017 | ***<br>0.52 | 0.043 | ****<br>0.75 |
| Field | | | | *<br>0.37 | 0.0097 | ***<br>0.57 |
| CT | | | | | 0.20 | 0.0069 |
| FA-II | | | | | | **<br>0.32 |

| Stroking | SA-II | HFA | Field | CT | FA-II | MS |
|---|---|---|---|---|---|---|
| SA-II | | 0.09 | 0.07 | ***<br>0.68 | **<br>0.6 | ****<br>0.76 |
| HFA | | | 0.0018 | ***<br>0.63 | ***<br>0.57 | ****<br>0.71 |
| Field | | | | **<br>0.47 | *<br>0.42 | ***<br>0.55 |
| CT | | | | | 0.066 | ***<br>0.58 |
| FA-II | | | | | | ***<br>0.6 |

| Holding | SA-II | HFA | Field | CT | FA-II | MS |
|---|---|---|---|---|---|---|
| SA-II | | *<br>0.34 | *<br>0.47 | **<br>0.54 | **<br>0.71 | ****<br>0.75 |
| HFA | | | 0.0098 | *<br>0.056 | *<br>0.31 | **<br>0.36 |
| Field | | | | 0.043 | **<br>0.49 | ***<br>0.5 |
| CT | | | | | *<br>0.46 | **<br>0.43 |
| FA-II | | | | | | 0.012 |

*Figure S1.* (A) Partial η² effect sizes for the LMEM tests reported in Figure 3A, 3B. (B) LMEM test results (upper line in each cell) and the corresponding partial η² effect sizes (bottom line in each cell) for the pairwise comparison reported in Figure 3C. *$p < 0.05$, **$p < 0.01$, ***$p < 0.001$, ****$p < 0.0001$.

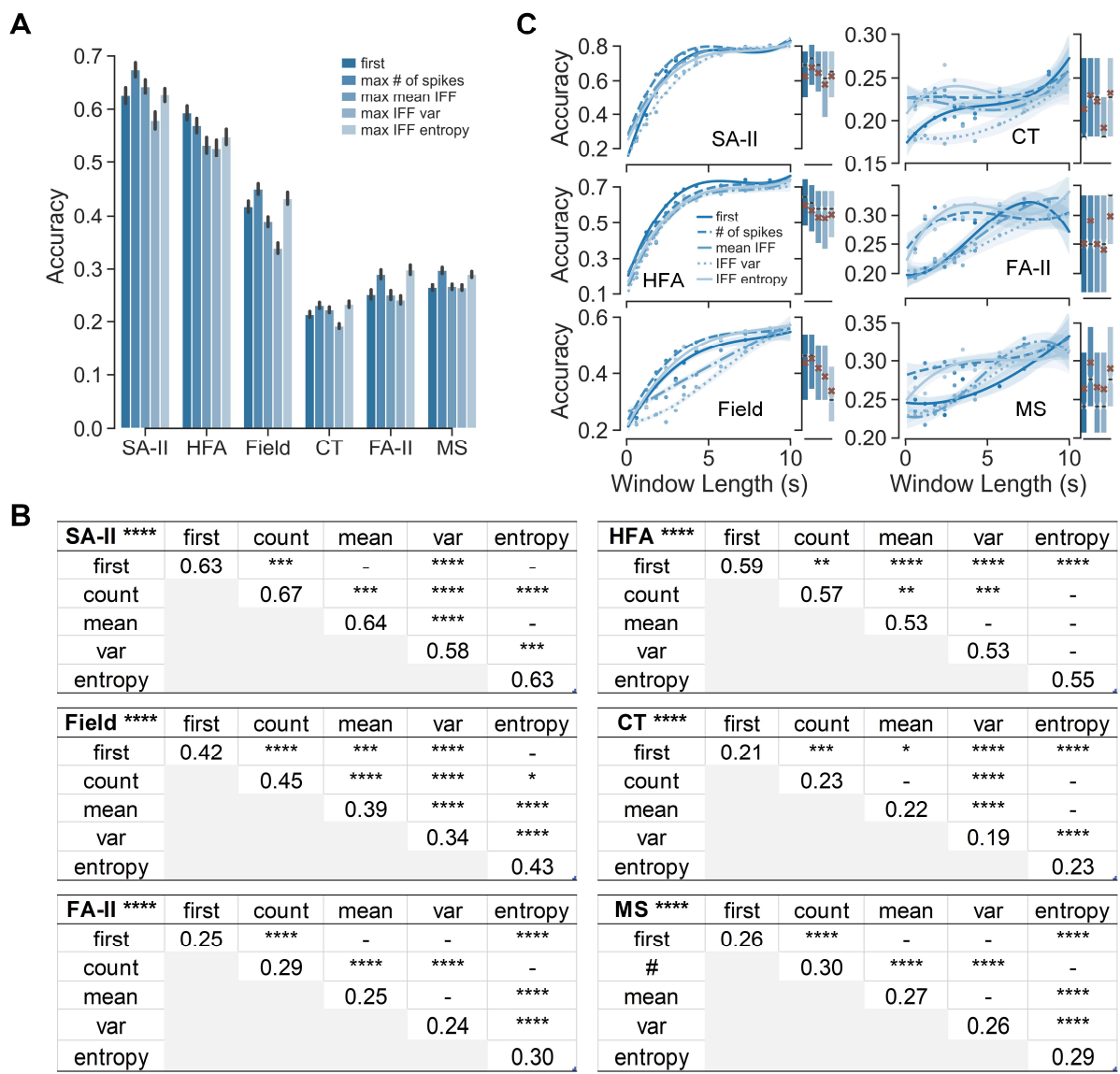

**B**

| SA-II **** | first | count | mean | var | entropy |
|---|---|---|---|---|---|
| first | 0.63 | *** | - | **** | - |
| count | | 0.67 | *** | **** | **** |
| mean | | | 0.64 | **** | **** |
| var | | | | 0.58 | *** |
| entropy | | | | | 0.63 |

| HFA **** | first | count | mean | var | entropy |
|---|---|---|---|---|---|
| first | 0.59 | ** | **** | **** | **** |
| count | | 0.57 | ** | *** | - |
| mean | | | 0.53 | - | - |
| var | | | | 0.53 | - |
| entropy | | | | | 0.55 |

| Field **** | first | count | mean | var | entropy |
|---|---|---|---|---|---|
| first | 0.42 | **** | *** | **** | - |
| count | | 0.45 | **** | **** | * |
| mean | | | 0.39 | **** | **** |
| var | | | | 0.34 | **** |
| entropy | | | | | 0.43 |

| CT **** | first | count | mean | var | entropy |
|---|---|---|---|---|---|
| first | 0.21 | *** | * | **** | **** |
| count | | 0.23 | - | **** | **** |
| mean | | | 0.22 | **** | **** |
| var | | | | 0.19 | **** |
| entropy | | | | | 0.23 |

| FA-II **** | first | count | mean | var | entropy |
|---|---|---|---|---|---|
| first | 0.25 | **** | **** | **** | **** |
| count | | 0.29 | **** | **** | - |
| mean | | | 0.25 | - | **** |
| var | | | | 0.24 | **** |
| entropy | | | | | 0.30 |

| MS **** | first | count | mean | var | entropy |
|---|---|---|---|---|---|
| first | 0.26 | **** | **** | **** | **** |
| # | | 0.30 | **** | **** | - |
| mean | | | 0.27 | - | **** |
| var | | | | 0.26 | **** |
| entropy | | | | | 0.29 |

*Figure S2.* (A) Classification accuracies across window position metrics averaged over all window lengths for each afferent subtype. (B) *p < 0.05, **p < 0.01, ***p < 0.001, ****p < 0.0001 were derived by Mann–Whitney U tests with Benjamini-Hochberg post-hoc correction. (C) Classification accuracies across window position metrics along with the increase of window length. Curves were fitted using third-order polynomial functions, points denote means of 10 evenly-binned data. Bar plots show distributions of classification accuracies over all window lengths per window position metric, and brown cross markers denote means per window position metric.